%% file: report.tex
\documentclass[journal]{IEEEtran}

\usepackage{cite}
\usepackage{stfloats}

\usepackage{mathtools}
\usepackage{array}
\usepackage{siunitx}

\usepackage{algpseudocode}
\usepackage{graphicx}
\graphicspath{{figs/}}
\DeclareGraphicsExtensions{.pdf,.jpeg,.png}
\ifCLASSOPTIONcompsoc
  \usepackage[caption=false,font=normalsize,labelfont=sf,textfont=sf]{subfig}
\else
  \usepackage[caption=false,font=footnotesize]{subfig}
\fi

\newcommand{\pso}{Particle Swarm Optimization}
\newcommand{\tsup}{\textsuperscript}

\newcommand\CONDITION[2]%
  {\begin{tabular}[t]{@{}l@{}l@{}}
     #1&#2
   \end{tabular}%
  }
\algdef{SE}[IF]{If}{EndIf}[1]%
  {\algorithmicif\ \CONDITION{#1}{\ \algorithmicthen}}%
  {\algorithmicend\ \algorithmicif}%

\DeclareMathOperator*{\argmin}{arg\,min}
\DeclareMathOperator*{\argmax}{arg\,max}
\DeclareMathOperator*{\sgn}{sgn}
\DeclareMathOperator*{\pbest}{\mathit{pbest}}
\DeclareMathOperator*{\gbest}{\mathit{gbest}}
\DeclareMathOperator*{\oldbest}{\mathit{oldbest}}
\DeclareMathOperator*{\lbest}{\mathit{lbest}}

\begin{document}

\title{A Novel Exoplanetary Habitability Score via Particle Swarm Optimization of CES Production Functions}


\author{Abhijit~Theophilus,~\IEEEmembership{Member,~IEEE,}
        Snehanshu~Saha,~\IEEEmembership{Senior Member,~IEEE,}
        Suryoday~Basak,~\IEEEmembership{Member,~IEEE
        and~Jayant~Murthy,~\IEEEmembership{Member,~IEEE}}
\thanks{}
\thanks{}
\thanks{Manuscript received ; revised }}

\markboth{ }%
{Theopolis \MakeLowercase{\textit{et. al.}}: }


\maketitle

\begin{abstract}
The search for life has two goals essentially: looking for
planets with Earth-like conditions (Earth similarity)
and looking for the possibility of life in some form
(habitability). Determining habitability from exoplanet data requires that determining parameters are collectively considered before coming up with a conclusion as no single factor alone contributes to it. Our proposed models, would serve as an indicator while looking for new habitable worlds, if computed with precision and efficiency. The models are of the type constrained optimization, multivariate, convex but may suffer from curvature violation and premature convergence impacting desired habitability scores. We mitigate the problem by proposing modified Particle Swarm Optimization (PSO) to tackle constraints and ensuring global optima. In the process, a python library to tackle such problems has been created.
\end{abstract}


\section{Introduction}\label{sec:intro}

\IEEEPARstart{T}{he} search for extra-terrestrial life~\cite{Seth1,Biosig} and potentially habitable extrasolar planets~\cite{Kepler,Presto} has been an international venture since Frank Drake's attempt with Project Ozma in the mid-20th century~\cite{Ozma}. Cochran, Hatzes, and Hancock\cite{Exo1} confirmed the first exoplanet in 1991. This marked the start of a trend that has lasted 25 years and yielded over 3,700 confirmed exoplanets. There have been attempts to assess the habitability of these planets and to assign a score based on their similarity to Earth. Two such habitability scores are the Cobb-Douglas Habitability (CDH) score~\cite{Bora,Saha} and the Constant Elasticity Earth Similarity Approach (CEESA) score. Estimating these scores involves maximizing a production function while observing a set of constraints on the input variables.

Under most paradigms, maximizing a continuous function requires calculating a gradient. This may not always be feasible for non-polynomial functions in high-dimensional search spaces. Further, subjecting the input variables to constraints, as needed by CDH and CEESA, are not always straightforward to represent within the model. This paper details an approach to Constrained Optimization (CO) using the swarm intelligence metaheuristic. \pso\ (PSO) is a method for optimizing a continuous function that does away with the need for calculating the gradient. It employs a large number of randomly initialized particles that traverse the search space, eventually converging at a global best solution encountered by at least one particle~\cite{PSO1,PSO2}.

\pso\ is a distributed method that requires simple mathematical operators and short segments of code, making it a lucrative solution where computational resources are at a premium. Its implementation is highly parallelizable. It scales with the dimensionality of the search space. The standard PSO algorithm does not deal with constraints but, through variations in initializing and updating particles, constraints are straightforward to represent and adhere to, as seen in Section~\ref{subsec:copso}. Poli\cite{Poli1,Poli2} carried out extensive surveys on the applications of PSO, reporting uses in Communication Networks, Machine Learning, Design, Combinatorial Optimization and Modeling, among others.

This paper demonstrates the applicability of \pso\ in estimating habitability scores, CDHS and CEESA of an exoplanet by maximizing their respective production functions (discussed in Sections~\ref{subsec:cdhs} and~\ref{subsec:ceesa}). CDHS considers the planet's Radius, Mass, Escape Velocity and Surface Temperature, while CEESA includes a fifth parameter, the Orbital Eccentricity of the planet. The Exoplanet Catalog hosted by the Planetary Habitability Laboratory, UPR Arecibo records these parameters for each exoplanet in Earth Units~\cite{PHL}. Section~\ref{sec:results} reports the performance of PSO and discusses the distribution of the habitability scores of the exoplanets.

\section{Habitability Scores}\label{sec:habscore}

\subsection{Cobb-Douglas Habitability Score}\label{subsec:cdhs}
Estimating the Cobb-Douglas Habitability (CDH) score~\cite{Bora} requires estimating an interior score ($\mathit{CDHS}_i$) and a surface score ($\mathit{CDHS}_s$) by maximizing the following production functions,
\begin{IEEEeqnarray}{llcrr}
  \IEEEyesnumber\IEEEyessubnumber*
  Y_i\ &=\ {CDHS}_i\ &=&\ R^\alpha.&D^\beta\,,\label{eq:cdhi}\\
  Y_s\ &=\ {CDHS}_s\ &=&\ {V_e}^\gamma.&{T_s}^\delta\,,\label{eq:cdhs}
\end{IEEEeqnarray}
where, $R$, $D$, $V_e$ and $T_s$ are density, radius, escape velocity and surface temperature respectively. $\alpha$, $\beta$, $\gamma$ and $\delta$ are the elasticity coefficients subject to $0 < \alpha,\beta,\gamma,\delta < 1$. Equations~\ref{eq:cdhi} and~\ref{eq:cdhs} are convex under either Constant Returns to Scale (CRS), when $\alpha+\beta=1$ and $\gamma+\delta=1$, or Decreasing Returns to Scale (DRS), when $\alpha+\beta<1$ and $\gamma+\delta<1$. The final CDH score is the convex combination of the interior and surface scores as given by,
\begin{equation}
  Y\ =\ w_i.Y_i + w_s.Y_s\,.
\end{equation}

\subsection{Constant Elasticity Earth Similarity Approach Score}\label{subsec:ceesa}
The Constant Elasticity Earth Similarity Approach (CEESA) uses the following production function to estimate the habitability score of an exoplanet,
\begin{equation}\label{eq:ceesa}
  Y = {(r.R^\rho+d.D^\rho+t.{T_s}^\rho+v.{V_e}^\rho+e.E^\rho)}^{\frac{\eta}{\rho}}\,,
\end{equation}
where, $E$ is the fifth parameter denoting Orbital Eccentricity. The value of $\rho$ lies within $0<\rho\leq 1$. The coefficients $r$, $d$, $t$, $v$ and $e$ lie in $(0,1)$ and sum to \num{1}, $r+d+t+v+e = 1$. The value of $\eta$ is constrained by the scale of production used, $0 < \eta < 1$ under DRS and $\eta=1$ under CRS.

\section{Particle Swarm Optimization}\label{sec:pso}

Particle Swarm Optimization (PSO)~\cite{PSO1} is a biologically inspired metaheuristic for finding the global minima of a function. Traditionally designed for unconstrained inputs, it works by iteratively converging a population of randomly initialized solutions, called particles, toward a globally optimal solution. Each particle in the population keeps track of its current position and the best solution it has encountered, called $\pbest$. Each particle also has an associated velocity used to traverse the search space. The swarm keeps track of the overall best solution, called $\gbest$. Each iteration of the swarm updates the velocity of the particle towards its $\pbest$ and the $\gbest$ values.

\subsection{PSO for Unconstrained Optimization}\label{subsec:uopso}
Let $f(x)$ be the function to be minimized, where $x$ is a $d$-dimensional vector. $f(x)$ is also called the fitness function. Algorithm~\ref{alg:unop} outlines the approach to minimizing $f(x)$ using PSO.\@ A set of particles are randomly initialized with a position and a velocity, where $l$ and $u$ are the lower and upper boundaries of the search space. The position of the particle corresponds to its associated solution. The algorithm initializes each particle's $\pbest$ to its initial position. The $\pbest$ position that corresponds to the minimum fitness is selected to be the $\gbest$ position of the swarm. Shi and Eberhart\cite{PSO2} discussed the use of inertial weights to regulate velocity to balance the global and local search. Upper and lower bounds limit velocity within $\pm v_\mathit{max}$.

On each iteration, the algorithm updates the velocity and position of each particle. For each particle, it picks two random numbers $u_g, u_p$ from a uniform distribution, $U(0,1)$ and updates the particle velocity. Here, $\omega$ is the inertial weight and $\lambda_g,\lambda_p$ are the global and particle learning rates. If the new position of the particle corresponds to a better fit than its $\pbest$, the algorithm updates $\pbest$ to the new position. Once the algorithm has updated all particles, it updates $\gbest$ to the new overall best position. A suitable termination criteria for the swarm, under convex optimization, is to terminate when $\gbest$ has not changed by the end of an iteration.

\begin{figure}[!t]
  \begin{algorithmic}[1]
    \Require{$f(x)$, the function to minimize.}
    \Ensure{global minimum of $f(x)$.}
    \For{each particle $i\gets 1,n$}
    \State{$p_i \sim {U(l,u)}^d$}
    \State{$v_i \sim {U(-|u-l|, |u-l|)}^d$}
      \State{${\pbest}_i \gets p_i$}
    \EndFor\
    \State{$\gbest \gets \smashoperator{\argmin\limits_{{\pbest}_i,\,i=1\dots n}} f({\pbest}_i)$}
    \Repeat\
      \State{$\oldbest \gets \gbest$}
      \For{each particle $i \gets 1\dots n$}
        \State{$u_p, u_g \sim U(0,1)$}
        \State{$v_i \gets \omega.v_i + \lambda_g u_g ({\gbest-p_i}) + \lambda_p u_p ({{\pbest}_i-p_i})$}
        \State{$v_i \gets \sgn(v_i).\max\{|v_\mathit{max}|, |v_i|\}$}
        \State{$p_i \gets p_i + v_i$}
        \If{$f(p_i) < f({\pbest}_i)$}
          \State{${\pbest}_i \gets p_i$}
        \EndIf\
      \EndFor\
      \State{$\gbest \gets \smashoperator{\argmin\limits_{{\pbest}_i,\,i=1\dots n}} f({\pbest}_i)$}
    \Until{$|\oldbest - \gbest| < \mathit{threshold}$}
    \State{\textbf{return} {$f(\gbest)$}}
  \end{algorithmic}
  \caption{Algorithm for PSO.}\label{alg:unop}
\end{figure}

\subsection{PSO with Leaders for Constrained Optimization}\label{subsec:copso}
Although PSO has eliminated the need to estimate the gradient of a function, as seen in Section~\ref{subsec:uopso}, it still is not suitable for constrained optimization. The standard PSO algorithm does not ensure that the initial solutions are feasible, and neither does it guarantee that the individual solutions will converge to a feasible global solution. Solving the initialization problem is straightforward, resample each random solution from the uniform distribution until every initial solution is feasible. To solve the convergence problem each particle uses another particle's $\pbest$ value, called $\lbest$, instead of its own to update its velocity. Algorithm~\ref{alg:cop} describes this process.

On each iteration, for each particle, the algorithm first picks two random numbers $u_g,u_p$. It then selects a $\pbest$ value from all particles in the swarm that is closest to the position of the particle being updated as its $\lbest$. The $\lbest$ value substitutes ${\pbest}_i$ in the velocity update equation. While updating $\pbest$ for the particle, the algorithm checks if the current fit is better than $\pbest$, and performs the update if the current position satisfies all constraints. The algorithm updates $\gbest$ as before.

\begin{figure}[!t]
  \begin{algorithmic}[1]
    \Require{$f(x)$, the function to minimize.}
    \Ensure{global minimum of $f(x)$.}
    \For{each particle $i\gets 1,n$}
      \Repeat\
      \State{$p_i \sim {U(l,u)}^d$}
      \Until{$p_i$ satisfies all constraints}
      \State{$v_i \sim {U(-|u-l|, |u-l|)}^d$}
      \State{${\pbest}_i \gets p_i$}
    \EndFor\
    \State{$\gbest \gets \smashoperator{\argmin\limits_{{\pbest}_i,\,i=1\dots n}} f({\pbest}_i)$}
    \Repeat\
      \State{$\oldbest \gets \gbest$}
      \For{each particle $i \gets 1\dots n$}
        \State{$u_p, u_g \sim U(0,1)$}
        \State{$\lbest \gets \smashoperator{\argmin\limits_{{\pbest}_j,\,j=1\dots n}} {\|{\pbest}_j -
            p_i\|}^2$}
        \State{$v_i \gets \omega.v_i + \lambda_g u_g ({\gbest-p_i}) + \lambda_p u_p ({\lbest-p_i})$}
        \State{$p_i \gets p_i + v_i$}
        \If{$f(p_i) < f({\pbest}_i)$ \textbf{and}\\$p_i$ satisfies all constraints}
          \State{${\pbest}_i \gets p_i$}
        \EndIf\
      \EndFor\
      \State{$\gbest \gets \smashoperator{\argmin\limits_{{\pbest}_i,\,i=1\dots n}} f({\pbest}_i)$}
    \Until{$|\oldbest - \gbest| < \mathit{threshold}$}
    \State{\textbf{return} {$f(\gbest)$}}
  \end{algorithmic}
  \caption{Algorithm for CO by PSO.}\label{alg:cop}
\end{figure}

\section{Representing the Problem}\label{sec:rep}

A Constrained Optimization problem can represented as,
\begin{IEEEeqnarray}{r'l'r}
  \IEEEnonumber*
  \underset{x}{\text{minimize}} & f(x)\\
  \text{subject to} & g_k(x)\leq 0, &\; k = 1\dots q\,,\\
                    & h_l(x) = 0,   &\; l = 1\dots r\,.
\end{IEEEeqnarray}

Ray and Liew\cite{Cop} describe a way to represent non-strict inequality constraints when optimizing using a particle swarm. Strict inequalities and equality constraints need to be converted to non-strict inequalities before being represented in the problem. Introducing an error threshold $\epsilon$ converts strict inequalities of the form ${g_k}^{\prime}(x) < 0$ to non-strict inequalities of the form $g_k(x) = {g_k}^{\prime}(x) + \epsilon \leq 0$. A tolerance $\tau$ is used to transform equality constraints to a pair of inequalities,
\begin{IEEEeqnarray}{rCr'l}
  \IEEEnonumber*
  g_{(q+l)}(x)   &=& h_l(x) - \tau \leq 0,	  & l = 1\dots r\,,\\
  g_{(q+r+l)}(x) &=& {-}h_l(x) - \tau \leq 0,& l = 1\dots r\,.
\end{IEEEeqnarray}

Thus, $r$ equality constraints become $2r$ inequality constraints, raising the total number of constraints to $s = q + 2r$. For each solution $p_i$, $c_i$ denotes the constraint vector where, $c_{ik} = \max\{g_k(p_i), 0\},~k=1\dots s$. When $c_{ik} = 0,~\forall k=1\dots s$, the solution $p_i$ lies within the feasible region. When $c_{ik} > 0$, the solution $p_i$ violates the $k$\tsup{th} constraint.

\subsection{Representing CDH Score Estimation}
Under the aforementioned guidelines, the representation of CDH score estimation under CRS is,
\begin{IEEEeqnarray}{r'r}\label{eq:cdhscrs}
  \IEEEnonumber
  \underset{\alpha,\beta,\gamma,\delta}{\text{minimize}}
      & \IEEEeqnarraymulticol{1}{l}{Y_i = {-}R^\alpha.D^\beta\,,}\\[-\belowdisplayskip]
      & \IEEEeqnarraymulticol{1}{l}{Y_s = {-}{V_e}^\gamma.{T_s}^\delta\,,}\\
  \IEEEyessubnumber*
  \text{subject to}
      &  {-}\phi + \epsilon\leq 0,\ \forall \phi\in\{\alpha,\beta,\gamma,\delta\}\,,\\
      & \phi - 1 + \epsilon\leq 0,\ \forall \phi\in\{\alpha,\beta,\gamma,\delta\}\,,\\
   	  &  \alpha+\beta-1 - \tau \leq 0\,,\label{con:c1}\\
	  &  1-\alpha-\beta - \tau \leq 0\,,\\
  	  & \gamma+\delta-1 - \tau \leq 0\,,\\
  	  & 1-\gamma-\delta - \tau \leq 0\,.\label{con:c2}
\end{IEEEeqnarray}
Under DRS the constraints~\ref{con:c1} to~\ref{con:c2} are replaced with,
\begin{IEEEeqnarray}{rCl}\label{eq:cdhsdrs}
	\IEEEyesnumber\IEEEyessubnumber*
     \alpha + \beta + \epsilon - 1 &\leq & 0\,,\\
    \gamma + \delta + \epsilon - 1 &\leq & 0\,.
\end{IEEEeqnarray}

\subsection{Representing CEESA}
The representation of CEESA score estimation (described in Section~\ref{subsec:ceesa}) under DRS is,

\begin{IEEEeqnarray}{r'rCl}\label{eq:ceesadrs}
  \IEEEnonumber
  \underset{r,d,t,v,e,\rho,\eta}{\text{minimize}}
  	& Y = {-}(r.R^\rho &+& d.D^\rho + t.{T_s}^\rho\\[-\belowdisplayskip]
    & &+& {v.{V_e}^\rho + e.E^\rho)} ^ {\frac{\eta}{\rho}}\\
  \IEEEyessubnumber*
  \text{subject to}
    & \IEEEeqnarraymulticol{3}{r}{\rho - 1 \leq 0\,,}\\
    & \IEEEeqnarraymulticol{3}{r}{\rho - 1 + \epsilon \leq 0\,,}\\
   	& \IEEEeqnarraymulticol{3}{r}{{-}\phi + \epsilon  \leq 0,\ \forall\phi\in\{r,d,t,v,e,\eta\}\,,}\\
    & \IEEEeqnarraymulticol{3}{r}{\phi - 1 + \epsilon \leq 0,\ \forall\phi\in\{r,d,t,v,e,\eta\}\,,}\\
    & \IEEEeqnarraymulticol{3}{r}{(r+d+t+v+e) - 1 - \tau \leq 0\,,}\\
    & \IEEEeqnarraymulticol{3}{r}{1 - (r-d-t-v-e) - \tau \leq 0\,.}
\end{IEEEeqnarray}
Under CRS there is no need for the parameter $\eta$ (since $\eta=1$). Thus, the objective function for the problem reduces to,
\begin{IEEEeqnarray}{r'rCl}\label{eq:ceesacrs}
  \IEEEnonumber
  \underset{r,d,t,v,e,\rho,\eta}{\text{minimize}}
  	& Y = {-}(r.R^\rho &+& d.D^\rho + t.{T_s}^\rho\\[-\belowdisplayskip]
    & &+& {v.{V_e}^\rho + e.E^\rho)} ^ {\frac{1}{\rho}}
\end{IEEEeqnarray}

\section{Experiment and Results}\label{sec:results}

\begin{table}[!t]
  \renewcommand{\arraystretch}{1.3}
  \caption{Parameters from the PHL-EC used for the experiment.}
  \label{tab:param}
  \centering
  \begin{tabular}{l l l}
    \hline
    \textbf{Parameter} & \textbf{Description} & \textbf{Unit}\\
    \hline
    P. Radius       & Estimated radius         & Earth Units (EU)\\
    P. Density      & Density                  & Earth Units (EU)\\
    P. Esc Vel      & Escape velocity          & Earth Units (EU)\\
    P. Ts Mean      & Mean Surface temperature & Kelvin (K)\\
    P. Eccentricity & Orbital eccentricity\\
    \hline
  \end{tabular}
\end{table}

\begin{figure*}[!t]
  \centering
  \subfloat[CRS Score Distribution]{%
  	\includegraphics[width=0.5\textwidth]{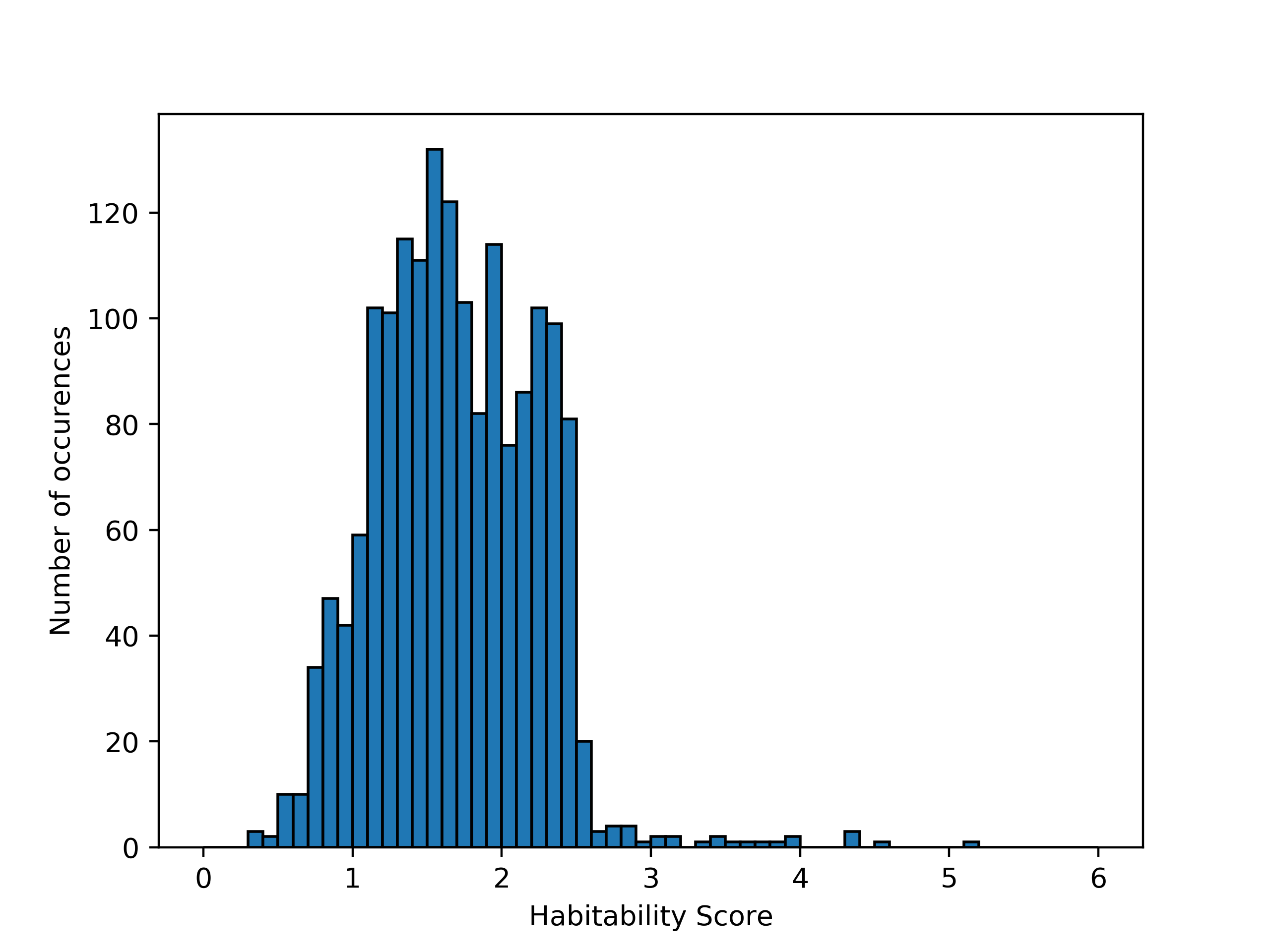}%
    \label{fig:distcdcrs}}
  \hfill
  \subfloat[DRS Score Distribution]{%
  	\includegraphics[width=0.5\textwidth]{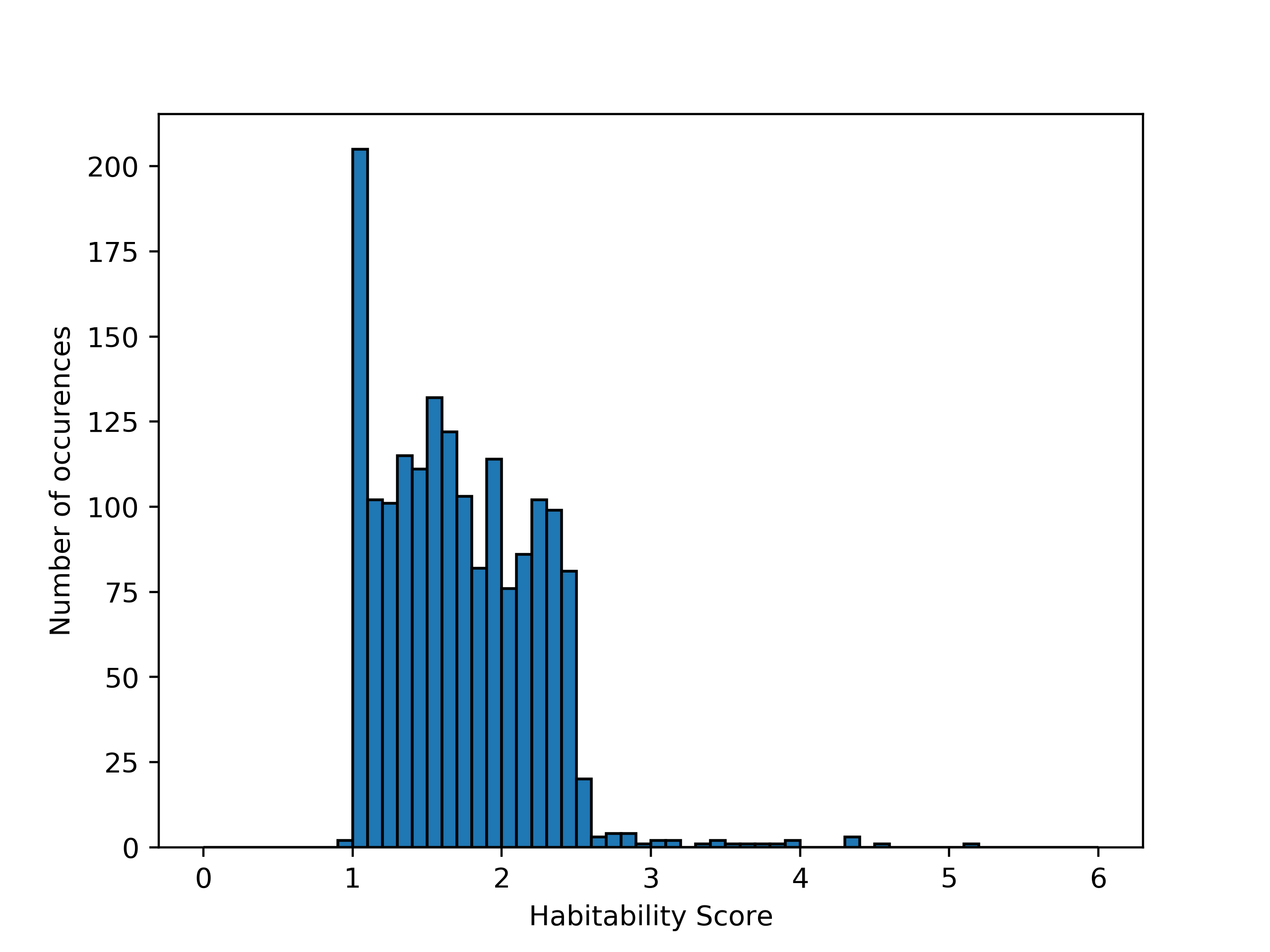}%
    \label{fig:distcddrs}}

  \subfloat[CRS Iterations Distribution]{%
  	\includegraphics[width=0.5\textwidth]{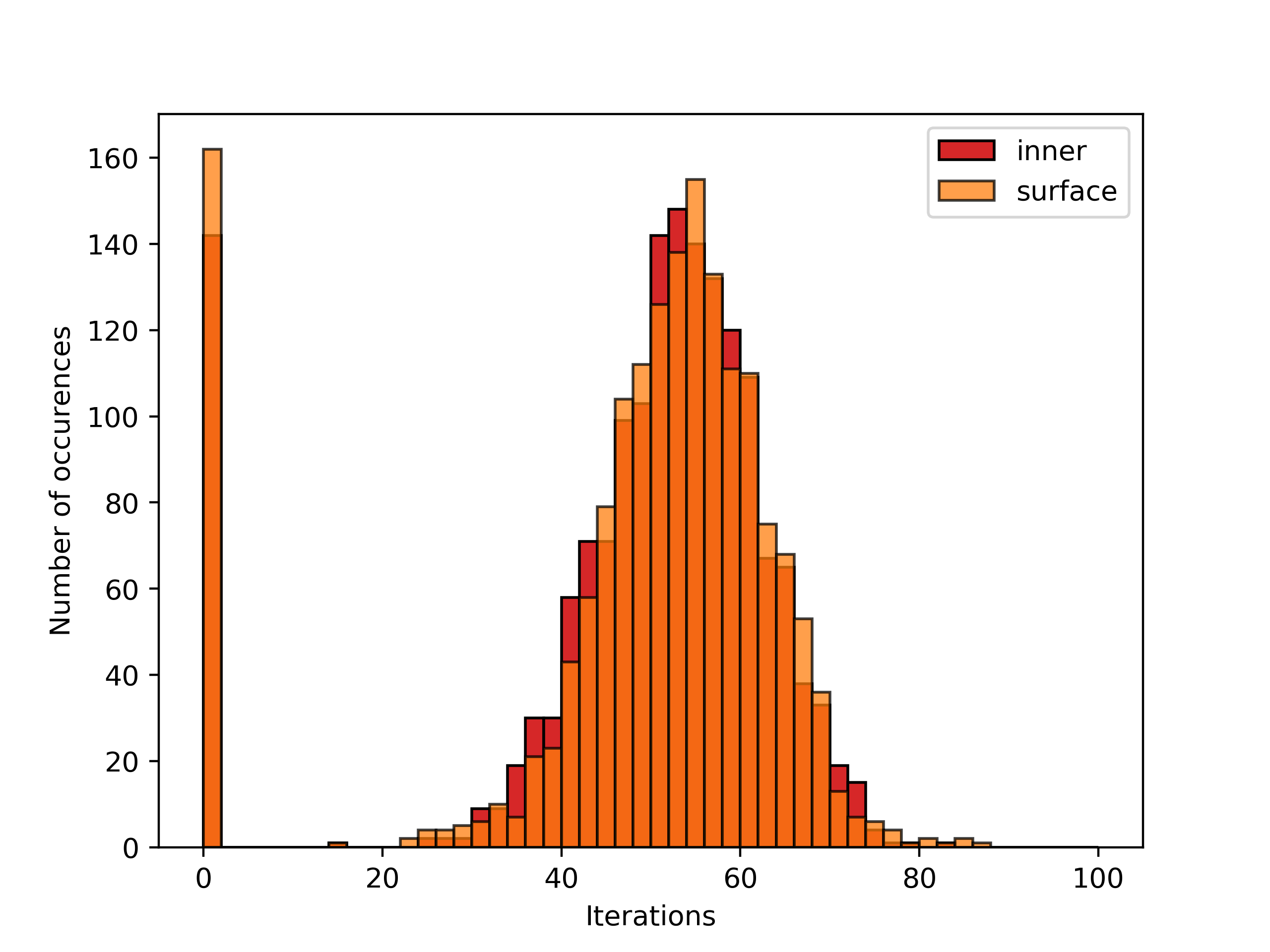}%
    \label{fig:itercdcrs}}
  \hfill
  \subfloat[DRS Iterations Distribution]{%
  	\includegraphics[width=0.5\textwidth]{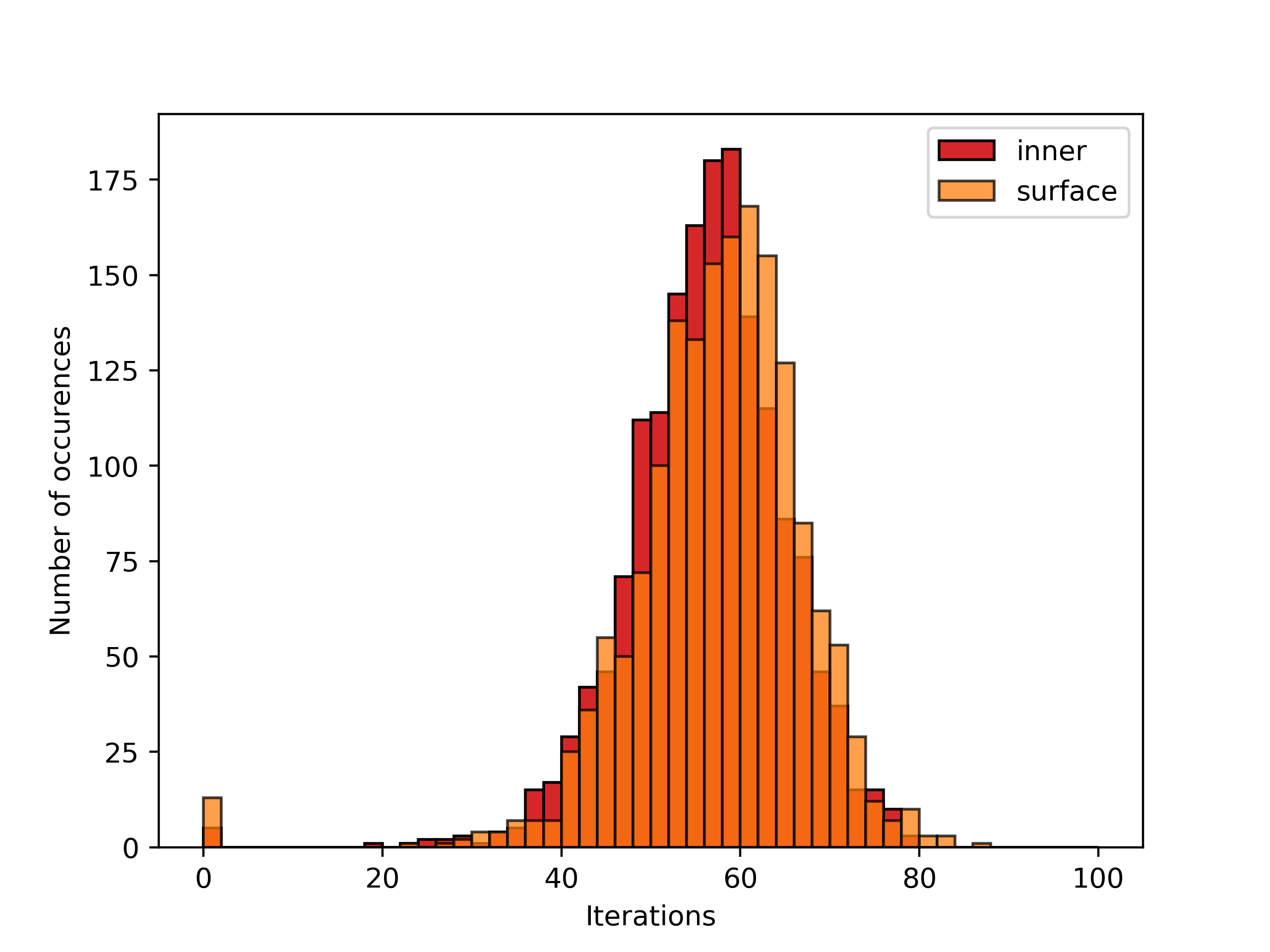}%
    \label{fig:itercddrs}}
  \caption{Plots for the Cobb-Douglas Habitability Score.}
  \label{fig:cdhs}
\end{figure*}

\begin{figure*}[!t]
  \centering
  \subfloat[CRS Score Distribution]{%
  	\includegraphics[width=0.5\textwidth]{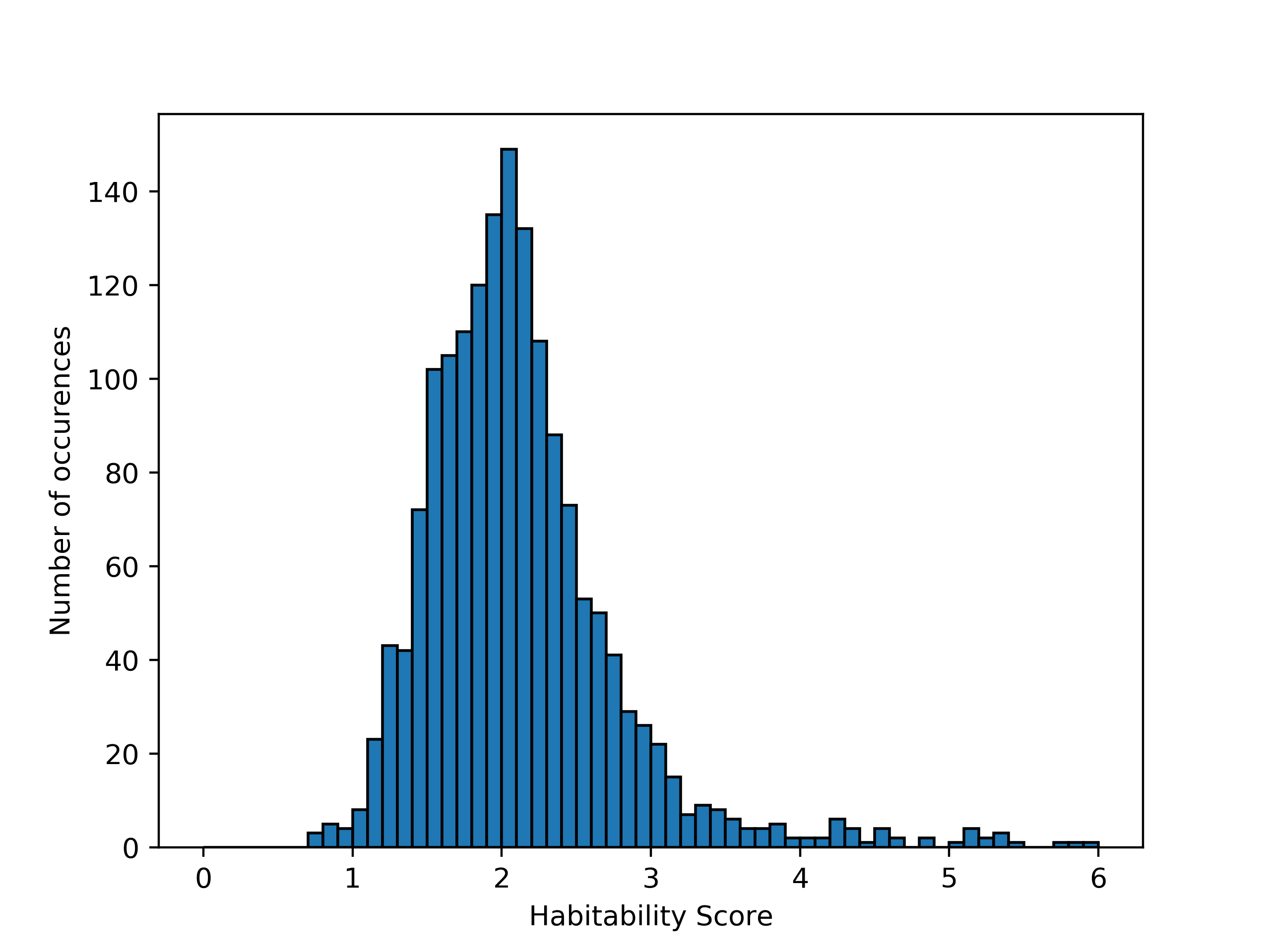}%
    \label{fig:distcecrs}}
  \hfill
  \subfloat[DRS Score Distribution]{%
  	\includegraphics[width=0.5\textwidth]{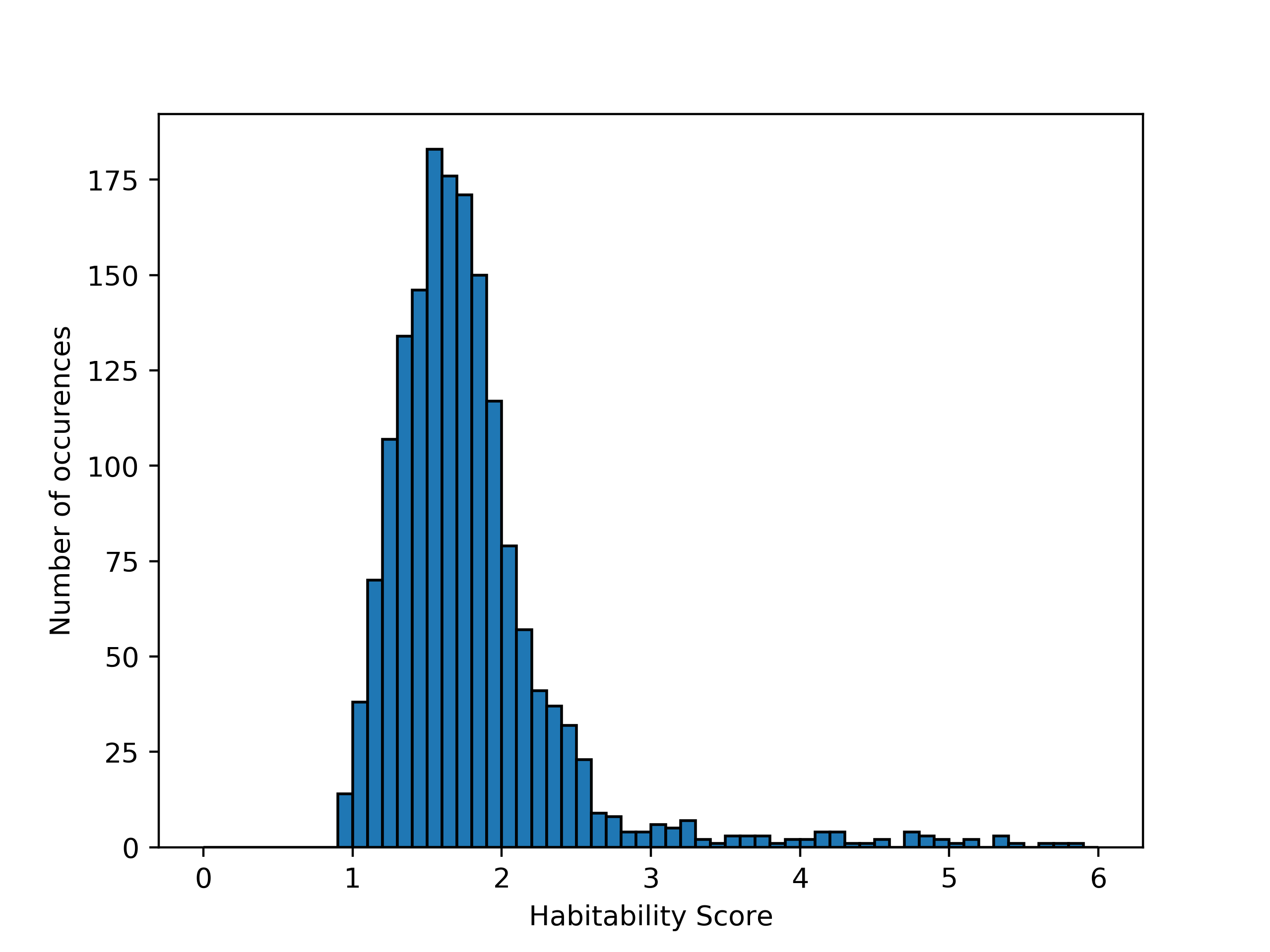}%
    \label{fig:distcedrs}}

  \subfloat[CRS Iterations Distribution]{%
  	\includegraphics[width=0.5\textwidth]{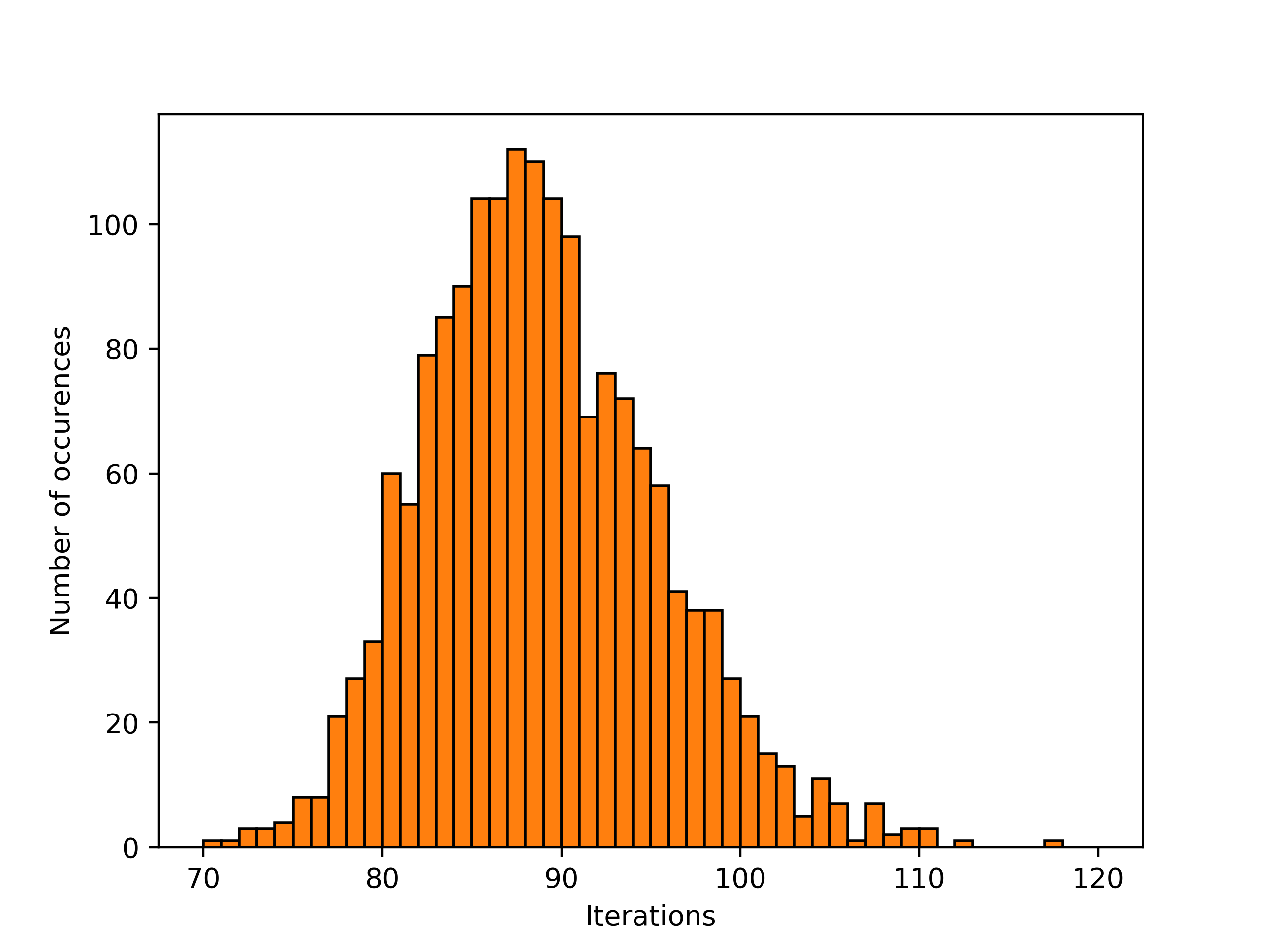}%
    \label{fig:itercecrs}}
  \hfill
  \subfloat[DRS Iterations Distribution]{%
  	\includegraphics[width=0.5\textwidth]{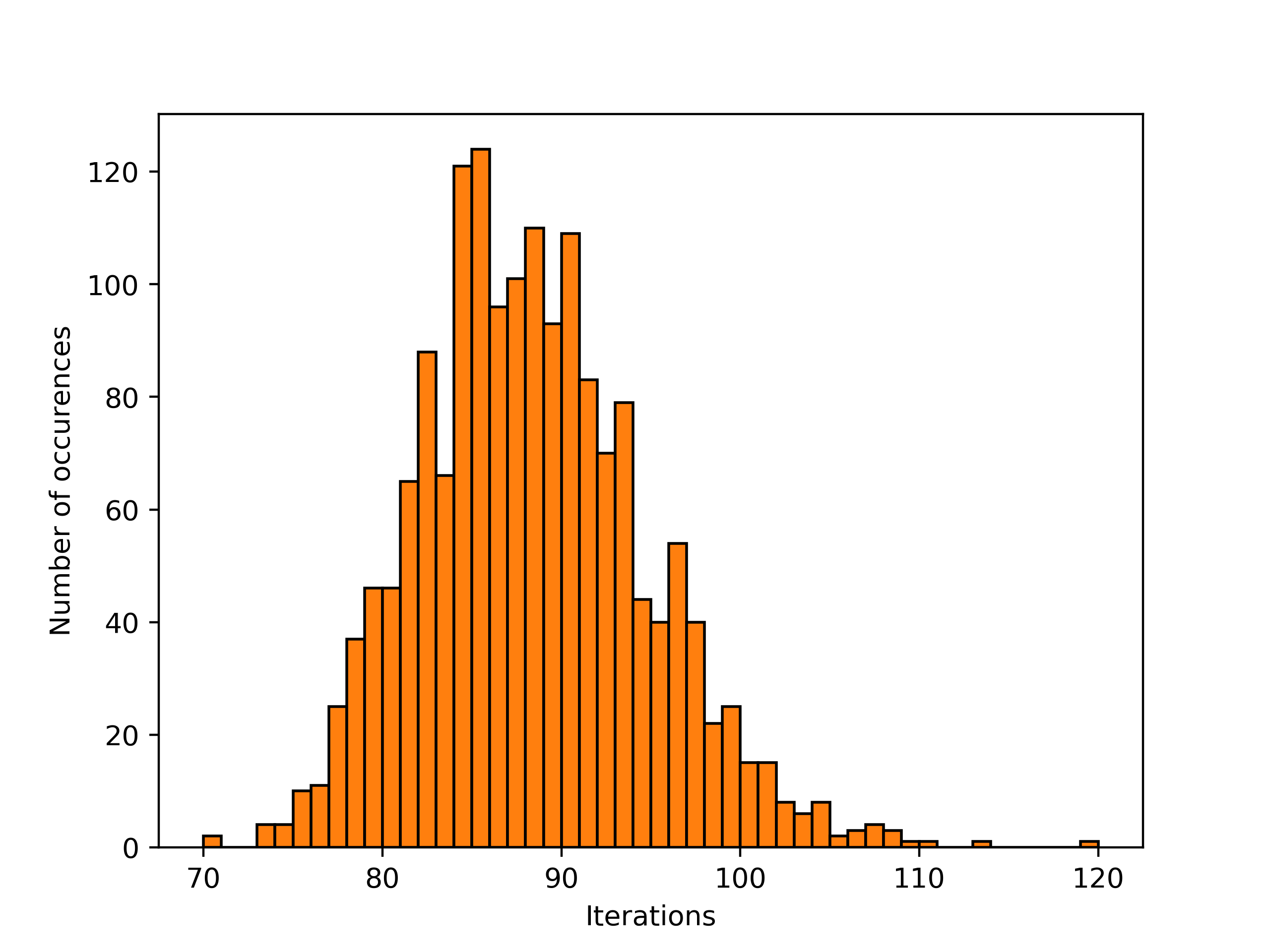}%
    \label{fig:itercedrs}}
  \caption{Plots for the Constant Elasticity Earth Similarity Approach.}
  \label{fig:ceesa}
\end{figure*}

The data set used for estimating the Habitability Scores of exoplanets was the Confirmed Exoplanets Catalog maintained by the Planetary Habitability Laboratory (PHL)~\cite{PHL}. The catalog records observed and modeled parameters for exoplanets confirmed by the Extrasolar Planets Encyclopedia. Table~\ref{tab:param} describes the parameters from the PHL Exoplanets' Catalog (PHL-EC) used for the experiment. Since surface temperature and eccentricity are not recorded in Earth Units, we normalized these values by dividing them with Earth's surface temperature (\SI{288}{\kelvin}) and eccentricity (\num{0.017}). PHL-EC assumes an Eccentricity of 0 when unavailable. The PHL-EC records empty values for planets whose surface temperature is not known. We chose to drop these records from the experiment.

The implementation resulted in a Python library now available on the Python Packaging Index under the name PSOPy~\cite{psopy}. It can be installed through pip as \texttt{pip install psopy} (also available in the github folder AstrIRG). Our implementation used $n=25$ particles to traverse the search space, with learning rates $\lambda_g=0.8$ and $\lambda_p=0.2$. It used an inertial weight of $\omega=0.6$ and upper and lower bounds $\pm1.0$. We used an error threshold of $\epsilon=\num{1e-6}$ to convert strict inequalities to non-strict inequalities, and a tolerance of $\tau=\num{1e-7}$ to transform an equality constraint to a pair of inequalities. Further implementation details are discussed in the Appendix.

The plots in Figures~\ref{fig:distcdcrs} and~\ref{fig:distcddrs} describe the distribution of the CDH scores across exoplanets tested from the PHL-EC.\@ Figures~\ref{fig:itercdcrs} and~\ref{fig:itercddrs} show the distribution of iterations required to converge to a global maxima. The spike at 0 is caused by particles converging to a $\gbest$ that does not shift from the original position (for a more detailed explanation see Appendix~\ref{app:con}). The plots in Figures~\ref{fig:distcedrs} and~\ref{fig:distcedrs} describe the distribution of the CEESA score across the exoplanets, while Figures~\ref{fig:itercedrs} and~\ref{fig:itercecrs} show the distribution of iterations to convergence. These graphs aggregate the results of optimizing the Habitability Production Functions (Equations~\ref{eq:cdhscrs},~\ref{eq:cdhsdrs},~\ref{eq:ceesadrs} and~\ref{eq:ceesacrs}) for each exoplanet in the PHL-EC by the method described in Algorithm~\ref{alg:cop}.

Table~\ref{tab:cdhscrs} records the CDH scores for a sample of exoplanets under CRS at $w_i=0.99$ and $w_s=0.01$. $\alpha$, $\beta$, $\gamma$ and $\delta$ record the parameters of Equation~\ref{eq:cdhscrs}. $Y_i$ and $Y_s$     record the maxima for the objective functions. $i_i$ and $i_s$ specify the number of iterations taken to converge to     a stable $\gbest$ value. Under the Class column there are four categories for the planets --- Psychroplanets (psy),     Mesoplanets (mes), Non-Habitable planets (non) and Hypopsychroplanets (hyp). Table~\ref{tab:cdhsdrs} records the CDH scores for a sample under DRS, with $\alpha$, $\beta$, $\gamma$ and $\delta$ recording the parameters of Equation~\ref{eq:cdhscrs}.

Tables~\ref{tab:ceesacrs} and~\ref{tab:ceesadrs} record the estimated CEESA scores under CRS and DRS respectively. $r$, $d$, $t$, $v$, $e$, $\rho$ and $eta$ record the parameters of Equation~\ref{eq:ceesadrs} in Table~\ref{tab:ceesadrs} and the parameters of Equation~\ref{eq:ceesacrs} in Table~\ref{tab:ceesacrs}. However, since under the CRS constraint, $\eta=1$, there is no need for the parameter $\eta$ in Table~\ref{tab:ceesacrs}. $i$ specifies the number of iterations taken to converge to the maxima.

These tables indicate that although CEESA has 7 parameters and 16 constraints under DRS, PSO takes a little over twice the number of iterations to converge as in each step of the CDH score estimation, which has 2 parameters and 5 constraints. This is a promising result as it indicates that the iterations required for converging increases sub-linearly with the number of parameters in the model. As for real time taken to converge, PSO took \SI{666.85}{\second} ($\approx\SI{11}{\minute}\ \SI{7}{\second}$) to estimate the CDH score under CRS for \num{1683} exoplanets, at an average of \SI{198.11}{\milli\second} for each planet for each individual score (interior and surface) of the CDH score.\@ For CDH estimation under DRS, it took \SI{638.69}{\second} ($\approx\SI{10}{\minute}\ \SI{39}{\second}$) at an average of \SI{189.75}{\milli\second} for each part of the CDH score.\@ The CEESA calculations, requiring a single estimate, took a little over half the CDH estimation execution time to run. Under DRS it took a total of \SI{370.86}{\second} ($\approx\SI{6}{\minute}\ \SI{11}{\second}$) at \SI{220.36}{\milli\second} per planet, while under CRS it took \SI{356.92}{\second} ($\approx\SI{5}{\minute}\ \SI{57}{\second}$) at \SI{212.07}{\milli\second} per planet.

\begin{table*}
  \renewcommand{\arraystretch}{1.3}
  \caption{Estimated Cobb-Douglas Habitability scores under CRS.}
  \label{tab:cdhscrs}
  \centering
  \include{./tabs/cdhscrs}
\end{table*}

\begin{table*}
  \renewcommand{\arraystretch}{1.3}
  \caption{Estimated Cobb-Douglas Habitability scores under DRS.}
  \label{tab:cdhsdrs}
  \centering
  \include{./tabs/cdhsdrs}
\end{table*}

\begin{table*}
  \renewcommand{\arraystretch}{1.3}
  \caption{Estimated Constant Elasticity Earth Similarity Approach scores under CRS.}
  \label{tab:ceesacrs}
  \centering
  \include{./tabs/ceesacrs}

\end{table*}

\begin{table*}
  \renewcommand{\arraystretch}{1.3}
  \caption{Estimated Constant Elasticity Earth Similarity Approach scores under DRS.}
  \label{tab:ceesadrs}
  \centering
  \include{./tabs/ceesadrs}

\end{table*}

\section{Conclusion}\label{sec:conc}

\pso\ mainly draws its advantages from being easy to implement and highly parallelizable. The algorithms described in Section~\ref{sec:pso} use simple operators and straightforward logic. What is especially noticeable is the lack of the need for a gradient, allowing PSO to work in high dimensional search spaces with a large number of constraints, precisely what is needed in a potential Habitability score estimate. Further, particles of the swarm, in most implementations operate independently during each iteration, their updates can occur simultaneously and even asynchronously, yielding much faster execution times than those outlined in Section~\ref{sec:results}. However, since strict inequalities and equality constraints are not exactly represented, the resulting solution may not be as accurate as direct methods. Despite this, using PSO to calculate the habitability scores is beneficial when the number of input parameters are large, which further increases the number of constraints, resulting in a model too infeasible for traditional optimization methods.
\par Determining habitability from exoplanet requires that determining parameters are collectively considered~\cite{PHL} before coming up with a conclusion as no single factor alone contributes to it. Our proposed model would serve as an indicator while looking for new habitable worlds. Eccentricity may have some effect on habitabilty and the models for computation should address that. CDHS doesn't, at least for the Trappist system (otherwise considered a set of potentially habitable exoplanets) since the eccentricities for all memebers of the Trappist system are 0 identically. CDHS, being a product model then will render the habitability score to be 0, not in agreement with observations and overall opinion in the community. This is the reason CESSA is considered in the process of habitability score computation (additive nature of the model). One might wonder why metaheuristic optimization was applied on two different optimization problems. We hope, our clarification would suffice.
\par However, the functional forms considered to compute the habitability score pose challenges. As we intend to add more parameters (such as eccentricity) to the basic model~\cite{PHL}\cite{Bora}, the functional form tends to suffer from curvature violation~\cite{saha2016novel}\cite{sarkar2016cdsfa}. Even though global optima is guaranteed, premature convergence and local oscillations are hard to mitigate. An attempt to address such issues, with moderate success, could be found in \cite{agrawal2018comparative}. The greatest contribution of the manuscript is to propose an evolutionary algorithm to track dynamic functions of the type that allow for the oscillation that were instead mitigated with SGA in \cite{sarkar2016cdsfa}. Consequently, a Pyhton library is integrated with the open source tool suite, an add on for coding enthusiasts to test our method.
\appendices
\section{Ensuring Convergence}\label{app:con}

Although the termination criteria illustrated in Algorithm~\ref{alg:unop}, $|\oldbest-\gbest| < \mathit{threshold}$, is a lucrative solution for unconstrained optimization, it might cause early termination for a constrained problem. This could occur in Algorithm~\ref{alg:cop} as the particles traverse the search space. Due to the way they are updated, all individual particles could move to infeasible regions. There are no updates made to their corresponding $\pbest$s, leading to no change in $\gbest$. Thus, although the swarm has not yet converged to the global optima, it terminates and reports the current $\gbest$ as the optima.

To prevent this from happening our implementation maintained a counter that it incremented every time $\gbest$ changed by a value less than the threshold. When the counter reached \num{100}, the algorithm terminated and reported the current value of $\gbest$ as the global optima. If the value of $\gbest$ changed by more than the threshold, it reset the counter to $\num{0}$. Since the function is convex, and all points are initialized to feasible solutions (thus, ensuring that a feasible $\gbest$ has been encountered) there was no need to restrict the total number of possible iterations.

This explains the spike at \num{0} in Figure~\ref{fig:itercdcrs}. For every exoplanet, the implementation waits a \num{100} iterations for declaring convergence, thus overstating the total iterations by \num{100}. To adjust for this offset, we subtracted \num{100} from the total iterations before constructing the histogram plots in Figures~\ref{fig:itercdcrs},~\ref{fig:itercddrs},~\ref{fig:itercecrs} and~\ref{fig:itercedrs}. Therefore, although the graphs report iterations to convergence for a significant number of planets, only $\gbest$ has not shifted, the particles have actually converged to a common point around a region within the threshold of the global best.

\section{Improving Performance}\label{app:imp}

Most programming languages today have either built-in or external library support for linear algebra routines. The implementation of these libraries are well-tested and efficient, and take advantage of system features such as multiple cores or a general purpose GPU computing device. Taking advantage of these libraries requires representing most entities as either matrices or vectors. 

Matrices $P$ and $V$ represent current position and velocity of all particles in the swarm.
\begin{IEEEeqnarray}{rCl:C:r}
  \IEEEnonumber*
  P &=& [\ p_{ij} &\mid &\forall i=1\dots n,\;\forall j=1\dots d\ ]\,,\\
  V &=& [\ v_{ij} &\mid &\forall i=1\dots n,\;\forall j=1\dots d\ ]\,.
\end{IEEEeqnarray}
$p_{ij}$ and $v_{ij}$ denote position and velocity of particle $i$ in dimension $d$. Matrices $B$ and $L$ represent the $\pbest$ and $\lbest$ values for the swarm, where,
\begin{IEEEeqnarray}{rCl}
  \IEEEnonumber*
  B &=& [\ {\pbest}_i\ \mid\ \forall i=1\dots n\ ]\,,\\
  L &=& [\ \argmin_{B_j,\ j = 1\dots n} {\|B_j - P_i\|}^2\ \mid\ \forall i=1\dots n\ ]\,.
\end{IEEEeqnarray}
Section~\ref{sec:rep} outlined the use of a constraint vector for describing the feasibility of a solution. This concept generalizes to a matrix $C$, where,
\begin{equation*}
  C = [\ c_{ij}\ \mid\ \forall i=1\dots n,\;\forall j=1\dots s\ ]\,.
\end{equation*}
Let $r',r''$ be two random vectors of length $n$ drawn from the uniform distribution ${U(0,1)}^n$. We use the shorthand notation of $X_i$ to denote the $i$\tsup{th} row of some matrix $X$. Let $f(X_i)$ be the fitness function and $\mathit{constraints}(X)$ construct $C$ for solutions $X$. $\gbest - P$ denotes the matrix $(\gbest-P_1, \gbest-P_2, \dots, P_n)$ and $\circ$ denotes the Hadamard product. The update of each particle in Algorithm~\ref{alg:cop} is implemented as,
\begin{IEEEeqnarray}{rCl}
  \IEEEnonumber*
  V  &=& \omega.V + \lambda_g.r'\circ(\gbest - P) + \lambda_p.r''\circ(L - P)\\
  V  &=& [\ \sgn(v_{ij})*\max\{|v_{max}|,|v_{ij}|\}\ \mid\\
  	  && \qquad\forall i=1\dots n,\;\forall j=1\dots d\ ]\\
  P  &=& P + V\\
  C  &=& \mathit{constraints}(P)\\
  B  &=& [\ X_i\ \mid\ (\textstyle\sum_{j=1}^s c_{ij} = 0 \to\\
  	  && \qquad\qquad X_i = \argmax\{f(P_i), f(B_i)\})\ \wedge\\
      && \qquad(\textstyle\sum_{j=1}^s c_{ij} \neq 0 \to X_i = B_i),\ \forall i=1\dots n\ ]
\end{IEEEeqnarray}

Finally, after all particles are updated, $\gbest$ is updated according to,
\begin{equation*}
  \gbest = \argmax_{{B}_i} \{\ {f(B_i)}\ \mid\ \forall i=\dots n\ \}\,.
\end{equation*}

\section*{Acknowledgment}
The authors would like the Science and Engineering
Research Board (SERB)-Department of Science and
Technology (DST), Government of of India for supporting
this research. The project reference number is: SERB-EMR/
2016/005687.

\ifCLASSOPTIONcaptionsoff
  \newpage
\fi


\bibliographystyle{IEEEtran}
\bibliography{IEEEabrv,./report.bib}

%








\end{document}

%% file: tabs/cdhscrs.tex
\begin{tabular}{l r r r r r r r r r r}
  \hline
  Name & Class & $\alpha$ & $\beta$ & $Y_i$ & $i_i$ & $\gamma$ & $\delta$ & $Y_s$ & $i_s$ & $\mathit{CDHS}$\\
  \hline
  GJ 176 b & non & $0.460$ & $0.540$ & $1.90$ & $50$ & $0.107$ & $0.893$ & $2.11$ & $61$ & $1.90$\\
  GJ 667 C b & non & $0.423$ & $0.577$ & $1.71$ & $58$ & $0.692$ & $0.308$ & $1.81$ & $54$ & $1.71$\\
  GJ 667 C e & psy & $0.129$ & $0.871$ & $1.40$ & $50$ & $0.258$ & $0.742$ & $1.39$ & $55$ & $1.40$\\
  GJ 667 C f & psy & $0.534$ & $0.466$ & $1.40$ & $48$ & $0.865$ & $0.135$ & $1.39$ & $47$ & $1.40$\\
  GJ 3634 b & non & $0.409$ & $0.591$ & $1.89$ & $58$ & $0.724$ & $0.276$ & $2.09$ & $48$ & $1.89$\\
  HD 20794 c & non & $0.260$ & $0.740$ & $1.35$ & $50$ & $0.096$ & $0.904$ & $1.34$ & $58$ & $1.35$\\
  HD 40307 e & non & $0.168$ & $0.832$ & $1.50$ & $49$ & $0.636$ & $0.364$ & $1.53$ & $63$ & $1.50$\\
  HD 40307 f & non & $0.702$ & $0.298$ & $1.52$ & $68$ & $0.303$ & $0.697$ & $1.55$ & $45$ & $1.52$\\
  HD 40307 g & psy & $0.964$ & $0.036$ & $1.82$ & $51$ & $0.083$ & $0.917$ & $1.98$ & $55$ & $1.82$\\
  Kepler-186 f & hyp & $0.338$ & $0.662$ & $1.17$ & $50$ & $0.979$ & $0.021$ & $1.12$ & $40$ & $1.17$\\
  Proxima Cen b & psy & $0.515$ & $0.484$ & $1.12$ & $37$ & $0.755$ & $0.245$ & $1.07$ & $ 0$ & $1.12$\\
  TRAPPIST-1 b & non & $0.319$ & $0.681$ & $1.09$ & $ 0$ & $0.801$ & $0.199$ & $0.89$ & $ 0$ & $1.09$\\
  TRAPPIST-1 c & non & $0.465$ & $0.535$ & $1.06$ & $ 0$ & $0.935$ & $0.065$ & $1.14$ & $26$ & $1.06$\\
  TRAPPIST-1 d & mes & $0.635$ & $0.365$ & $0.77$ & $34$ & $0.475$ & $0.525$ & $0.73$ & $47$ & $0.77$\\
  TRAPPIST-1 e & psy & $0.145$ & $0.855$ & $0.92$ & $ 0$ & $0.897$ & $0.103$ & $0.83$ & $55$ & $0.92$\\
  TRAPPIST-1 g & hyp & $0.226$ & $0.774$ & $1.13$ & $43$ & $0.876$ & $0.124$ & $1.09$ & $ 0$ & $1.13$\\
  \hline
\end{tabular}

%% file: tabs/cdhsdrs.tex
\begin{tabular}{l r r r r r r r r r r}
  \hline
  Name & Class & $\alpha$ & $\beta$ & $Y_i$ & $i_i$ & $\gamma$ & $\delta$ & $Y_s$ & $i_s$ & $\mathit{CDHS}$\\
  \hline
  GJ 176 b & non & $0.395$ & $0.604$ & $1.90$ & $59$ & $0.372$ & $0.627$ & $2.11$ & $56$ & $1.90$\\
  GJ 667 C b & non & $0.781$ & $0.218$ & $1.71$ & $58$ & $0.902$ & $0.097$ & $1.81$ & $57$ & $1.71$\\
  GJ 667 C e & psy & $0.179$ & $0.820$ & $1.40$ & $49$ & $0.234$ & $0.765$ & $1.39$ & $60$ & $1.40$\\
  GJ 667 C f & psy & $0.704$ & $0.295$ & $1.40$ & $64$ & $0.398$ & $0.601$ & $1.39$ & $61$ & $1.40$\\
  GJ 3634 b & non & $0.602$ & $0.397$ & $1.89$ & $59$ & $0.429$ & $0.570$ & $2.09$ & $77$ & $1.89$\\
  HD 20794 c & non & $0.014$ & $0.985$ & $1.35$ & $50$ & $0.116$ & $0.883$ & $1.34$ & $45$ & $1.35$\\
  HD 40307 e & non & $0.752$ & $0.247$ & $1.50$ & $60$ & $0.677$ & $0.322$ & $1.53$ & $50$ & $1.50$\\
  HD 40307 f & non & $0.887$ & $0.112$ & $1.52$ & $51$ & $0.261$ & $0.738$ & $1.55$ & $60$ & $1.52$\\
  HD 40307 g & psy & $0.300$ & $0.699$ & $1.82$ & $62$ & $0.785$ & $0.214$ & $1.98$ & $56$ & $1.82$\\
  Kepler-186 f & hyp & $0.073$ & $0.926$ & $1.17$ & $46$ & $0.740$ & $0.259$ & $1.12$ & $51$ & $1.17$\\
  Proxima Cen b & psy & $0.045$ & $0.954$ & $1.12$ & $57$ & $0.216$ & $0.783$ & $1.07$ & $53$ & $1.12$\\
  TRAPPIST-1 b & non & $0.102$ & $0.897$ & $1.09$ & $41$ & $0.000$ & $0.000$ & $1.00$ & $65$ & $1.09$\\
  TRAPPIST-1 c & non & $0.471$ & $0.528$ & $1.06$ & $44$ & $0.227$ & $0.772$ & $1.14$ & $57$ & $1.06$\\
  TRAPPIST-1 d & mes & $0.000$ & $0.000$ & $1.00$ & $67$ & $0.000$ & $0.000$ & $1.00$ & $59$ & $1.00$\\
  TRAPPIST-1 e & psy & $0.000$ & $0.000$ & $1.00$ & $55$ & $0.000$ & $0.000$ & $1.00$ & $57$ & $1.00$\\
  TRAPPIST-1 g & hyp & $0.888$ & $0.111$ & $1.13$ & $47$ & $0.949$ & $0.050$ & $1.09$ & $46$ & $1.13$\\
  \hline
\end{tabular}

%% file: tabs/ceesacrs.tex
\begin{tabular}{l r r r r r r r r r r}
  \hline
  Name & Class & $r$ & $d$ & $t$ & $v$ & $e$ & $\rho$ & $\eta$ & $\mathit{CDHS}$ & $i$\\
  \hline
GJ 176 b & non & $0.194$ & $0.020$ & $0.315$ & $0.465$ & $0.006$ & $0.398$ & $1.000$ & $1.88$ & $ 86$\\
GJ 667 C b & non & $0.162$ & $0.289$ & $0.090$ & $0.087$ & $0.372$ & $0.836$ & $1.000$ & $3.54$ & $107$\\
GJ 667 C e & psy & $0.373$ & $0.032$ & $0.134$ & $0.304$ & $0.157$ & $0.217$ & $1.000$ & $1.25$ & $ 71$\\
GJ 667 C f & psy & $0.394$ & $0.006$ & $0.043$ & $0.360$ & $0.196$ & $0.490$ & $1.000$ & $1.44$ & $ 81$\\
GJ 3634 b & non & $0.351$ & $0.122$ & $0.006$ & $0.069$ & $0.453$ & $0.439$ & $1.000$ & $2.89$ & $ 96$\\
HD 20794 c & non & $0.101$ & $0.077$ & $0.691$ & $0.071$ & $0.059$ & $0.756$ & $1.000$ & $1.58$ & $ 94$\\
HD 40307 e & non & $0.069$ & $0.091$ & $0.097$ & $0.173$ & $0.569$ & $0.768$ & $1.000$ & $5.29$ & $ 94$\\
HD 40307 f & non & $0.285$ & $0.161$ & $0.053$ & $0.443$ & $0.058$ & $0.342$ & $1.000$ & $1.42$ & $ 73$\\
HD 40307 g & psy & $0.156$ & $0.010$ & $0.081$ & $0.302$ & $0.451$ & $0.612$ & $1.000$ & $7.15$ & $ 94$\\
Kepler-186 f & hyp & $0.036$ & $0.017$ & $0.082$ & $0.383$ & $0.483$ & $0.929$ & $1.000$ & $1.68$ & $ 85$\\
Proxima Cen b & psy & $0.352$ & $0.383$ & $0.103$ & $0.059$ & $0.103$ & $0.936$ & $1.000$ & $0.89$ & $ 83$\\
TRAPPIST-1 b & non & $0.148$ & $0.147$ & $0.344$ & $0.269$ & $0.093$ & $0.767$ & $1.000$ & $0.94$ & $ 81$\\
TRAPPIST-1 c & non & $0.038$ & $0.060$ & $0.575$ & $0.321$ & $0.005$ & $0.602$ & $1.000$ & $1.17$ & $ 86$\\
TRAPPIST-1 d & mes & $0.023$ & $0.065$ & $0.475$ & $0.391$ & $0.045$ & $0.830$ & $1.000$ & $0.84$ & $ 79$\\
TRAPPIST-1 e & psy & $0.176$ & $0.464$ & $0.253$ & $0.103$ & $0.004$ & $0.920$ & $1.000$ & $0.86$ & $ 81$\\
TRAPPIST-1 g & hyp & $0.060$ & $0.086$ & $0.310$ & $0.540$ & $0.004$ & $0.848$ & $1.000$ & $0.97$ & $ 86$\\
  \hline
\end{tabular}

%% file: tabs/ceesadrs.tex
\begin{tabular}{l r r r r r r r r r r}
  \hline
  Name & Class & $r$ & $d$ & $t$ & $v$ & $e$ & $\rho$ & $\eta$ & $\mathit{CDHS}$ & $i$\\
  \hline
GJ 176 b & non & $0.304$ & $0.001$ & $0.375$ & $0.271$ & $0.050$ & $0.467$ & $0.808$ & $1.52$ & $ 85$\\
GJ 667 C b & non & $0.297$ & $0.010$ & $0.318$ & $0.052$ & $0.322$ & $0.682$ & $0.730$ & $2.36$ & $ 90$\\
GJ 667 C e & psy & $0.230$ & $0.286$ & $0.137$ & $0.199$ & $0.148$ & $0.551$ & $0.906$ & $1.14$ & $ 85$\\
GJ 667 C f & psy & $0.397$ & $0.035$ & $0.152$ & $0.402$ & $0.014$ & $0.793$ & $0.999$ & $1.31$ & $100$\\
GJ 3634 b & non & $0.178$ & $0.175$ & $0.005$ & $0.194$ & $0.447$ & $0.894$ & $0.657$ & $2.07$ & $ 94$\\
HD 20794 c & non & $0.073$ & $0.142$ & $0.452$ & $0.190$ & $0.144$ & $0.953$ & $0.635$ & $1.20$ & $ 78$\\
HD 40307 e & non & $0.156$ & $0.307$ & $0.185$ & $0.033$ & $0.319$ & $0.428$ & $0.939$ & $2.69$ & $ 88$\\
HD 40307 f & non & $0.272$ & $0.231$ & $0.064$ & $0.305$ & $0.127$ & $0.676$ & $0.802$ & $1.28$ & $ 77$\\
HD 40307 g & psy & $0.113$ & $0.219$ & $0.066$ & $0.454$ & $0.148$ & $0.711$ & $0.991$ & $3.26$ & $ 92$\\
Kepler-186 f & hyp & $0.039$ & $0.159$ & $0.116$ & $0.329$ & $0.357$ & $0.253$ & $0.919$ & $1.35$ & $ 70$\\
Proxima Cen b & psy & $0.272$ & $0.173$ & $0.284$ & $0.193$ & $0.079$ & $0.615$ & $0.114$ & $0.99$ & $ 75$\\
TRAPPIST-1 b & non & $0.488$ & $0.151$ & $0.039$ & $0.193$ & $0.129$ & $0.151$ & $0.014$ & $0.99$ & $ 87$\\
TRAPPIST-1 c & non & $0.172$ & $0.236$ & $0.275$ & $0.242$ & $0.075$ & $0.969$ & $0.962$ & $1.06$ & $ 80$\\
TRAPPIST-1 d & mes & $0.106$ & $0.308$ & $0.075$ & $0.218$ & $0.293$ & $0.844$ & $0.017$ & $0.99$ & $ 93$\\
TRAPPIST-1 e & psy & $0.189$ & $0.266$ & $0.192$ & $0.094$ & $0.260$ & $0.371$ & $0.006$ & $0.99$ & $ 84$\\
TRAPPIST-1 g & hyp & $0.326$ & $0.186$ & $0.143$ & $0.278$ & $0.067$ & $0.315$ & $0.021$ & $1.00$ & $ 76$\\
  \hline
\end{tabular}